\documentclass[12pt]{article}
\usepackage{graphicx}
\usepackage{float}
\usepackage{subcaption}

\newcommand\pubnumber{NuPhys2018-Tenti}
\newcommand\pubdate{\today}

\def\Title#1{\begin{center} {\Large #1 } \end{center}}

\newcommand\pubblock{\rightline{\begin{tabular}{l} \pubnumber\\
         \pubdate  \end{tabular}}}
\newenvironment{Abstract}{\begin{quotation}  }{\end{quotation}}
\newenvironment{Presented}{\begin{quotation} \begin{center} 
             PRESENTED AT\end{center}\bigskip 
      \begin{center}\begin{large}}{\end{large}\end{center} \end{quotation}}
\def\Acknowledgements{\bigskip  \bigskip \begin{center} \begin{large}
             \bf ACKNOWLEDGEMENTS \end{large}\end{center}}

\def\beq{\begin{equation}}
\def\eeq#1{\label{#1}\end{equation}}
\def\eeqn{\end{equation}}

\def\beqa{\begin{eqnarray}}
\def\eeqa#1{\label{#1}\end{eqnarray}}
\def\eeqan{\end{eqnarray}}

\let\bar=\overbar

\def\Dslash{\not{\hbox{\kern-4pt $D$}}}
\def\dslash{\not{\hbox{\kern-2pt $\del$}}}

\def\msb{{\bar{\ssstyle M \kern -1pt S}}}

\begin{document}
\begin{titlepage}
\pubblock

\vfill
\Title{The ENUBET narrow band neutrino beam}
\vfill
\begin{center}
M.~Tenti\\
\vspace{.1cm}
{\small INFN, Sezione di Milano-Bicocca, Piazza della Scienza 3, Milano, Italy} \\ 
{\small Phys. Dep. Universit\`a di Milano-Bicocca, Piazza della scienza 3, Milano, Italy} \\ 
\vspace{.5cm}
On behalf of the ENUBET Collaboration:\\
\vspace{.1cm}
F.~Acerbi,
G.~Ballerini,
M.~Bonesini,
C.~Brizzolari,
G.~Brunetti
M.~Calviani,
S.~Carturan,
M.G.~Catanesi,
S.~Cecchini,
F.~Cindolo,
G.~Collazuol,
E.~Conti
F.~Dal~Corso,
G.~De Rosa,
C.~Delogu,
A.~Falcone,
B.~Goddard,
A.~Gola,
R.A.~Intonti,
C.~Jollet,
V.~Kain,
B.~Klicek,
Y.~Kudenko,
M.~Laveder,
A.~Longhin,
P.F.~Loverre,
L.~Ludovici,
L.~Magaletti,
G.~Mandrioli,
A.~Margotti,
V.~Mascagna,
N.~Mauri,
A.~Meregaglia,
M.~Mezzetto,
M.~Nessi,
A.~Paoloni,
M.~Pari,
E.~Parozzi,
L.~Pasqualini,
G.~Paternoster,
L.~Patrizii,
C.~Piemonte,
M.~Pozzato,
F.~Pupilli,
M.~Prest,
E.~Radicioni,
C.~Riccio,
A.C.~Ruggeri,
G.~Sirri,
M.~Soldani,
M.~Stipcevic,
F.~Terranova,
M.~Torti,
E.~Vallazza,
F.~Velotti,
M.~Vesco,
L.~Votano
\end{center}

\vfill
\begin{Abstract}
The narrow band beam of ENUBET is the first implementation of the ``monitored neutrino beam" technique proposed in 2015. 
ENUBET has been designed to monitor lepton production in the decay tunnel of neutrino beams and to provide a 1\% measurement of the neutrino flux at source. In particular, the three body semi-leptonic decay of kaons monitored by large angle positron production offers a fully controlled $\nu_{e}$ source at the GeV scale for a new generation of short baseline experiments. In this contribution the performances of the positron tagger prototypes tested at CERN beamlines in 2016-2018 are presented. 

\end{Abstract}
\vfill
\begin{Presented}
NuPhys2018, Prospects in Neutrino Physics
Cavendish Conference Centre, London, UK, December 19--21, 2018
\end{Presented}
\vfill
\end{titlepage}
\def\thefootnote{\fnsymbol{footnote}}
\setcounter{footnote}{0}

\section{Introduction}

The ENUBET (Enhanced NeUtrino BEams from kaon Tagging) project \cite{Longhin:2014yta,enubet} is aiming to demonstrate the feasibility of a monitored $\nu_{e}$ beam approach. The idea is based on the monitoring of the production of large angle positron in the decay tunnel from $K^{+} \rightarrow \pi^{0} e^{+} \nu_{e}$ ($K^{+}_{e3}$) decays. Thanks to the optimization of the focusing and transport system of the momentum-selected narrow band beam of the parent mesons, the $K^{+}_{e3}$ decay represents the main source of electron neutrino. Furthermore, the positron rate is a direct handle to measure the electron neutrino flux. Consequently, the monitored electron neutrino beam will lower the uncertainties on the neutrino flux from the current level of about $\mathcal{O}(7\% - 10\%)$ for conventional beam to $\sim 1\%$. 

A controlled neutrino source, like the one proposed by ENUBET, could be exploited by future experiments aiming at precise electron neutrino cross section measurements. It could be also exploited in a phase-II sterile neutrino search, especially in case of a positive signal from the upcoming short baseline experiments. Finally, ENUBET intends to set the first milestone towards a “time-tagged neutrino beam”, where the $\nu_{e}$ at the detector is time-correlated with the produced $e^{+}$ in the
decay tunnel \cite{Acerbi:2019qiv}.

\section{The positron tagger}
The positron monitoring is achieved instrumenting the decay tunnel. The requirements for the detector are: \textit{(i)} a high $e/\pi$ separation capability to remove the main source of background; \textit{(ii)} radiation hardness, since it will operate in a harsh environment; \textit{(iii)} cost effectiveness, since a large fraction of the $\sim$ 50 m long decay tunnel will be instrumented; \textit{(iv)} fast recovery time to cope with the expected rate of 200 kHz/cm$^{2}$. 

\begin{figure}[htb]
\centering
\includegraphics[width=0.5\textwidth]{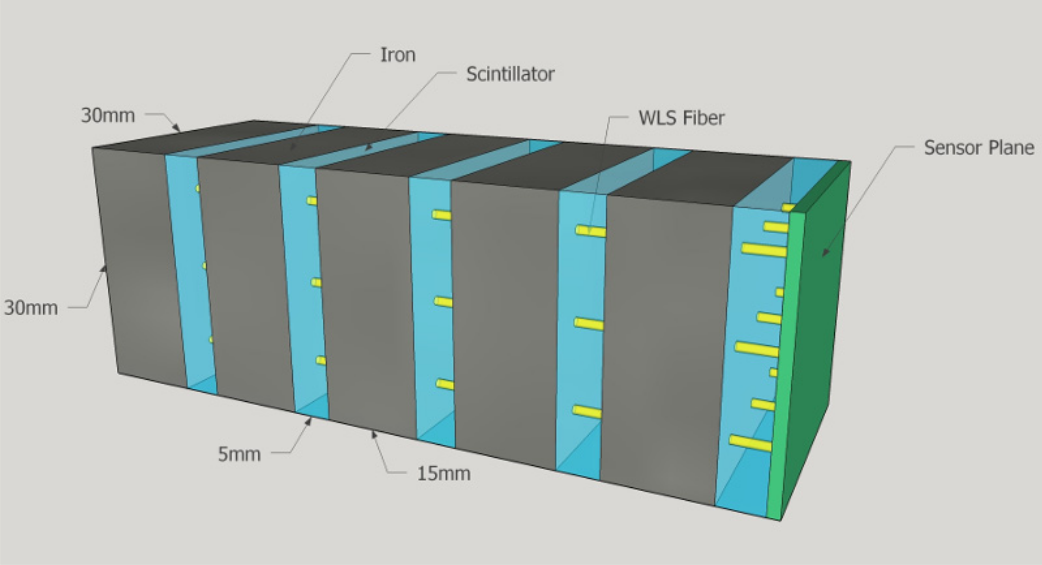}
\caption{Scheme of the baseline UCM. It is composed by five, 15 mm thick, iron layers interleaved by 5 mm thick plastic scintillator tiles. The total length of the module (10 cm) and its transverse size (3 $\times$ 3 cm$^{2}$) corresponds to 4.3 $X_{0}$ and 1.7 Moliere radii, respectively}
\label{fig:UCM}
\end{figure}

A shashlik calorimeter \cite{Fessler:1984wa,Atoian:1992ze} with longitudinal segmentation was identified as a solution  fulfilling all these requirements. The calorimeter basic unit (Fig.~\ref{fig:UCM}), the Ultra-Compact Module (UCM), is made of five, 15 mm thick, iron layers interleaved by 5 mm thick plastic scintillator tiles \cite{Berra:2016thx,Berra:2017rsi}. The total length of the module (10 cm) and its transverse size (3 $\times$ 3 cm$^{2}$) corresponds to 4.3 $X_{0}$ and 1.7 Moliere radii, respectively. Nine wavelength shifting (WLS) fibers crossing the UCM are connected directly to 1 mm$^{2}$ SiPMs through a plastic holder. SiPMs are hosted on a PCB embedded in the calorimeter structure. 

In November 2016, a prototype consisting of a 7 $\times$ 4 $\times$ 2 array of UCMs was exposed, at the CERN PS East Area facility, to beam composed by electrons, muons, and pions with momentum in the range of interest for neutrino physics applications $\left(1 - 5 \textrm{ GeV}\right)$, in order to measure its performances \cite{Ballerini:2018hus}. The scintillator tiles were machined and polished from EJ-200 and BC-412 sheets and painted with a diffusive TiO$_{2}$ based coating (EJ-510) to increase the light collection efficiency. After painting, nine holes with a diameter of 1.2 $\pm$ 0.1 mm were drilled in each tile with a CNC machine in order to accommodate the WLS fibers (Y11 and BCF92) read out by FBK 20 $\mu$m-pixel SiPMs based on the n-on-p RGB-HD technology. 

Ancillary detectors were used to identify particle type and isolate samples of electrons and muons. During the data taking the calorimeter was tilted at different angles (0, 50, 100, 200 mrad) with respect to the beam direction. The tilted geometry reproduces the operating condition of the calorimeter in the decay tunnel of neutrino beams, where positrons from $K^{+}_{e3}$ reach the detector with an average angle of $\sim$ 100 mrad. 

The measured deposited energy is in good agreement with the results obtained from a Monte Carlo simulation that took into account all the different components of the beam (namely electrons, muons and pions). The electromagnetic energy resolution is well described by $\sim 17\%/\sqrt{E/\textrm{GeV}}$ showing that the dominant contribution to the resolution is due to the sampling term and that the energy response is not affected by the light readout scheme employed to achieve longitudinal segmentation. A good $e/\pi$ separation, with a pion mis-identification lower than 3\%, is achieved exploiting the longitudinal segmentation of the calorimeter and the different energy deposit patterns. Furthermore, the analysis of the tilted runs indicates that the performance of the calorimeter is similar to the 0 mrad run in the angular range of interest for ENUBET. Finally, the calorimeter shows linear response, within $<$ 3\%, in the whole range of interest in both standard (0 mrad) and tilted runs.  

\section{Neutron irradiation test of RGB-HD SiPMs}
The integration of the light read out into the calorimeter module is a very effective solution but results into exposing the SiPMs to fast neutrons produced by hadronic showers. The non-ionizing fluence integrated during the lifetime of the experiment and scaled to 1 MeV equivalent neutrons is 1.8 $\times$ 10$^{11}$ n/cm$^{2}$.
In order to evaluate the effect of the radiation on the performances of the RGB-HD SiPMs, three PCBs hosting 9 SiPMs and the single-SiPM PCB were irradiated from a minimum dose of 1.8 $\times$ 10$^{8}$ n/cm$^{2}$ up to 1.7 $\times$ 10$^{11}$ n/cm$^{2}$ at the irradiation facility of INFN-LNL (Laboratori Nazionali di Legnaro) based on the CN Van Der Graaf accelerator \cite{Acerbi:2019wti}. 
All the SiPMs show minor changes in the breakdown voltage, while the dark current after breakdown increases by more than two orders of magnitude at a fluence of $\sim$ 10$^{11}$ n/cm$^{2}$. Finally, the sensitivity to single photoelectron is lost at fluences larger than 3 $\times$ 10$^{9}$ n/cm$^{2}$.

Later in October 2017, the irradiated PCBs were installed in two calorimeter prototypes and tested on the T9 beamline of the CERN East Area facility. A prototype (16B) had the layout descried in the previous section, while a second one (17UA) was built from injection molded scintillator tiles produced by Uniplast (Russia) for ENUBET. In the latter one, each tile is made by 3 extruded polystyrene based scintillator slabs (3 $\times$ 3 cm$^{2}$, 4.5 mm thickness) for a total thickness of 1.35 cm. 
In both, the sensitivity to mips was monitored since it is very important for calibration purposes during the whole life of the detector. 
The 16B prototype is not able to separate a mip from the noise peak up to the maximum fluence expected in ENUBET (2 $\cdot$ 10$^{11}$ n/cm$^{2}$) due to the increase of the dark counts, even if the electron peak remains well separated from noise.
On the other hand, for the 17UA prototype which employs the same SiPM-to-fiber coupling scheme as 16B but with a larger scintillator thickness, the mip peak remains separated from the dark noise peak even after irradiation. 
The electron and mip peak mean value ratio is constant after irradiation and the integrated neutron fluence does not affect the dynamic range of the photosensors. Hence, for the SiPMs employed in this test saturation effects of the signal due to the reduction of the number of working pixels after irradiation are not visible at $\mathcal{O}$(10$^{11}$ n/cm$^{2}$).
  

\section{The polysiloxane prototype}
Non conventional options based on polysiloxane scintillators are also being scrutinized. Polysiloxane is a recently developed siliconic-based scintillator \cite{polysiloxane} that offers several advantages over plastic scintillators: better radiation tolerance, reduced aging, no irreversible deterioration caused by mechanical deformations, exposure to solvent vapours and high temperatures and, for shashlik calorimeters, no need to drill and insert the optical fibers. Silicone rubbers preserve their transparency even after a 10 kGy dose exposure and their physical properties are constant over a wide temperature range. Polysiloxane can be poured at 60$^{\circ}$ C in the shashlik module after the insertion of the optical fibers in the absorber greatly simplifying the assembly. On the other hand, the light yield is about 30\% of the EJ-200 yield.

A polysiloxane shashlik calorimeter prototype composed of three modules (6 $\times$ 6 $\times$ 15 cm$^{3}$ each, 13 $X_{0}$ in total) consisting of 12 UCM units and a 15 mm scintillator thickness was built at INFN-LNL and exposed to particle beams at the CERN East area in October 2017. An electromagnetic energy resolution of $\sim$ 17\%/$\sqrt{E/\textrm{GeV}}$ and good linearity ($<$ 3\% in the 1-5 GeV range) were measured. Since, the light yield per unit thickness is about one-third of EJ-200, the quality of the fiber-scintillator coupling after pouring is comparable to the one that can be obtained from injection molding of conventional scintillator.

\section{Lateral scintillation light readout prototype}
In 2018, a different solution for the scintillation light read out was studied. Light is collected from both sides of each scintillator tile through WLS fiber accommodated in suitable groove machined on the plastic. Fibers from the same UCM are bundled to a single SiPM reading in groups of 10 at a distance of about 30 cm from the bulk of the calorimeter. In this layout SiPMs are not embedded to the calorimeter bulk, thus not exposed to the hadronic shower. This solution is less compact than a shashlik UCM but provides a safer environment for the SiPMs and allows easier access regarding maintenance or replacements. Furthermore, the lateral scheme is mechanically simpler both from the point of view of machining of the calorimeter component and for the fiber-to-SiPM coupling. 

In 2018, a prototype made of 3 $\times$ 2 $\times$ 2 UCMs was assembled. Each UCM was composed of 1.5 cm thick iron slabs interleaved with 0.5 cm plastic scintillators (EJ-204): the modules have thus the same sampling term as the shashlik UCM. The prototype was tested at CERN at the T9 beamline in May 2018. Later, in September, a larger prototype consisting of 84 UCMs in the 7 $\times$ 4 $\times$ 3 structure was studied at CERN. Preliminary results show an electromagnetic energy resolution at 1 GeV of about 17\% which is consistent with the prediction of 15\% obtained from a Monte Carlo simulation that does not include photon generation and transport. The analysis of this prototype is in progress.

\begin{figure}[htb]
\begin{subfigure}{0.5\textwidth}
\includegraphics[width=\textwidth]{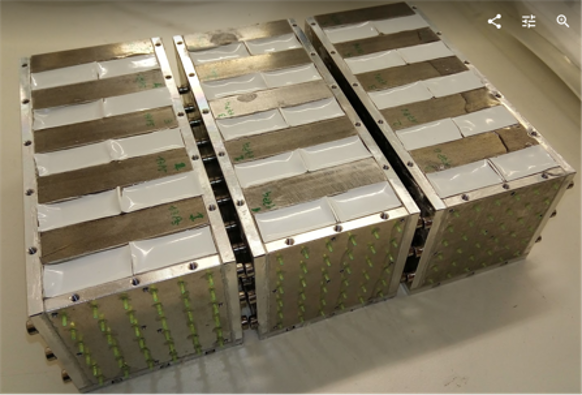}
\caption{}
\label{fig:16}
\end{subfigure} 
\begin{subfigure}{0.5\textwidth}
\includegraphics[width=\textwidth]{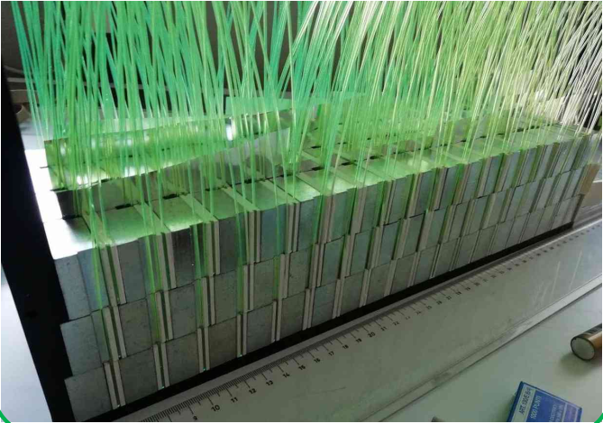}
\caption{}
\label{fig:17UA}
\end{subfigure}
\caption{Pictures of the polysiloxane-based (a) and lateral scintillation light readout (b) calorimeter prototypes.}
\label{fig:animals}
\end{figure}

\Acknowledgements
This project has received funding from the European Research Council (ERC) under the European Unions Horizon 2020 research and innovation programme (grant agreement N. 681647).

\end{document}